\RequirePackage{ifpdf}
\ifpdf 
\documentclass[pdftex]{sigma}
\else
\documentclass{sigma}
\fi

\def\ot{\otimes}

\def\kap{\kappa}

\def\ep{\epsilon}
\def\th{\theta}

\def\da{\dagger}
\def\De{\Delta}
\def\de{\delta}
\def\bet{\beta}
\def\ov{\over}
\def\ha{{1\over 2}}

\def\l{\left}
\def\l({\left(}
\def\r){\right)}
\def\r{\right}

\def\la{\lambda}
\def\al{\alpha}

\begin{document}

\numberwithin{equation}{section}

\allowdisplaybreaks

\renewcommand{\PaperNumber}{040}

\FirstPageHeading

\renewcommand{\thefootnote}{$\star$}

\ShortArticleName{$q$-Boson in Quantum Integrable Systems}

\ArticleName{$\boldsymbol{q}$-Boson in Quantum Integrable
Systems\footnote{This paper is a contribution to the Vadim
Kuznetsov Memorial Issue ``Integrable Systems and Related
Topics''. The full collection is available at
\href{http://www.emis.de/journals/SIGMA/kuznetsov.html}{http://www.emis.de/journals/SIGMA/kuznetsov.html}}}

\Author{Anjan KUNDU} \AuthorNameForHeading{A. Kundu}

\Address{Saha Institute of Nuclear Physics,
 Theory Group {\rm \&}
 Centre for Applied Mathematics \\ {\rm \&} Computational Science, 1/AF Bidhan Nagar, Calcutta 700 064, India}

\Email{\href{mailto:anjan.kundu@saha.ac.in}{anjan.kundu@saha.ac.in}}
\URLaddress{\url{http://www.saha.ac.in/theory/anjan.kundu/}}

\ArticleDates{Received November 14, 2006, in f\/inal form January
15, 2007; Published online March 05, 2007}

\Abstract{$q$-bosonic realization  of the underlying Yang--Baxter
algebra
 is identif\/ied for a~series of   quantum integrable systems, including some new models
like two-mode  $q$-bosonic model leading to a coupled
two-component
 derivative NLS model,
 wide range of $q$-deformed matter-radiation models, $q$-anyon model etc.
Result on a new exactly solvable interacting
 anyon gas,  linked to
 $q$-anyons on the lattice is  reported.}


\Classification{81R50; 81R12; 81T27; 81V80}

\section{Introduction}

In a quantum system the related  commutation rules determine the
character of  basic operators involved in the model, e.g.\
 boson, fermion, spin, anyon
etc. When such quantum systems are integrable
 the commutation rules are
determined by the underlying Yang--Baxter (YB)  algebra obtained
from the
 quantum Yang--Baxter equation (QYBE)
\begin{gather}
  R(\la-\mu)L_j(\la)\otimes L_j(\mu)=
(I \otimes L_j(\mu)) (L_j(\la)\otimes I)R(\la-\mu), \qquad j=1,2,
\dots ,N \label{qybe}
\end{gather}
together with the ultralocality condition
 $  L_j(\la)\otimes L_k(\mu)=
(I \otimes L_k(\mu)) (L_j(\la)\otimes I),$ at $k \not =j$, which
are at the same time suf\/f\/icient for the integrability of the
system.
 Here $L_j(\la)$ is the representative  Lax operator of the  lattice (or
discretized)  quantum  integrable systems   (QIS) at each lattice
site $ j=1,2, \dots ,N $ and  $R(\la-\mu) $ is the quantum
$R$-matrix
  ($c$-number matrix) which determines the structure constants of the YB algebra.
For rational solution of the $R$-matrix one usually gets the spin
algebra or its bosonic realization, while the trigonometric
solution yields $q$-spin and $q$-bosons, which are
 comparatively new entries in the f\/ield of quantum physics, discovered
mainly from the study of integrable systems \cite{qg,qg1,qg2}.

The concept of $q$-boson  was  formally introduced  as
$q$-parameter deformation of a standard boson, or through two-mode
realization of quantum algebra   \cite{qbose,qbose1,qbose2}.
Subsequently, the f\/in\-ding of   more application oriented
single mode $q$-bosonic realization of  quantum algebra through
$q$-de\-formation of Holstein--Primakov transformation
 \cite{qbs,qbs1,kunduq}
and  linking it to  QIS \cite{bogolq}
 was a~signif\/icant  step. However, it seems that this initial but
important step has not been pursued with enough zest to explore
the possibility of appearance
 of $q$-boson in  other QIS and
 its   app\-lications in exactly solvable physical models.
It is true that there were  attempts  to construct interesting
$q$-bosonic and $q$-spin
 models in various f\/ields and to study their physical ef\/fects
 \cite{qbs,qbs1,qappl,qappl1,qappl2,qappl3,qappl4}.
 However
  most of  such  models  could not be related to the underlying YB algebra
of a QIS permitting   exact treatment by the  Bethe ansatz (BA)
method.
 Our aim here therefore, is mainly to
 identify the $q$-bosonic mode  in a variety of QIS, which is deeply rooted
in  discrete or  lattice regularized models allowing  BA
solutions.
 We
discover also the appearance of such modes in some new QIS, e.g.\
 physically relevant
integrable matter-radiation models, a coupled double-mode
$q$-bosonic model leading to a two-component derivative NLS
 model etc.\
and  furthermore by  introducing  a   $q$-anyon like concept
f\/ind  its
  application  in another
 new exactly solvable derivative-$ \delta $ function  1D anyon gas.

\section[$q$-boson and $q$-spin]{$\boldsymbol{q}$-boson and $\boldsymbol{q}$-spin}

We review in brief the basic formulation of the $q$-boson and its
relation to various other objects like $q$-spin, standard boson,
canonical variables etc. $q$-boson  can be def\/ined through
a~deformation  of the bosonic commutation relation (CR):
\begin{gather}
  b_q   b_q^ \dag  -  q^{- 1} b_q^ \dag   b_q= q^{N}
, \qquad [b_q,N] = b_q, \qquad  [b_q^\dag,N] =- b_q^ \dag,
\label{qboson}
\end{gather}
which is supposed to be invariant under a ref\/lection $ q \to
q^{-1 }$. Therefore combining such two relations we can def\/ine
the $q$-boson in a more symmetric form
\begin{gather}
   [ b_q, b_q^ \dag ] = {
\cos (\al (2N+1)) \over \cos {\al } } , \qquad [b_q,N] = b_q,
\qquad  [b_q^\dag,N] =- b_q^ \dag, \label{qbos}
\end{gather}
where $ q=e^{2i \alpha}.$ At $q \to 1 $ the $q$-boson goes into  a
standard boson with
 both the above relations  reducing to
the bosonic CR:
\begin{gather}
  [ b, b^ \dag ] = 1
, \qquad [b,N] = b, \qquad  [b^\dag,N] =- b^ \dag. \label{bos}
\end{gather}
One can f\/ind also a direct mapping between $q$-deformed and
undeformed bosons
  as
\begin{gather}
 b_q=bf(N), \qquad b^\dag_q=f(N)b^\dag, \qquad  f(N)=
\left(\frac{[N]_{q}}{N }\right)^{\ha}, \qquad N=b^\dag b,
 \label{qb-b}
 \end{gather}
where $[x]_{q}= { q^x-q^{-x} \over q-q^{-1} }$. Using these
relations and the usual representation of bosonic operators
 it is easy therefore to construct a $q$-bosonic representation \cite{qbose,qbose1,qbose2}.
 Initially the $q$-boson was introduced as a  two-mode
realization  of the  $q$-spin ($q$-deformed Schwinger
transformation)
\begin{gather*}
    s^+_q= b^{\dag }_{ q 1}b_{ q 2}, \qquad  s^-_q= b^{\dag }_{ q 2}b_{ q 1}, \qquad
s^3=\ha(N_1-N_2)
 \end{gather*}
linked to quantum algebra
 $su_q(2)$
 \begin{gather}
    [s^+_q,  s^-_q]= \epsilon
[2s^3]_q,\qquad [s^3,  s^\pm_q]= \pm s^\pm_q,
 \label{qs}
 \end{gather}
 for $\epsilon =1 $.  $\epsilon =-1 $  on the other hand   gives
 noncompact algebra  $su_q(1,1)$.
 Subsequently, a single mode realization was
found through $q$-Holstein--Primakov ($q$-HP) transformation
\cite{qbs,qbs1,kunduq}
 \begin{gather}
 s^3=s \mp N, \qquad   s^+_q= g_\pm (N,q) b_q, \qquad  s_q^-= b_q^\dag g_\pm (N,q)
, \qquad g_{\pm}^{2}(N,q)=[2s \mp N]_q, \label{qHolst}
\end{gather}
for  $su_q(2) $ and  $su_q(1,1) $, respectively,
 with $ s$-being a spin parameter.
One can f\/ind also mapping from $q$-boson  to canonical
variables: $[u,p ]=i $ as
\begin{gather}
 b_q=e^{-ip}f_c(u), \qquad b^\dag_q=f_c(u) e^{ip}, \qquad   f^2_c(u)=[c-u]_q
 \label{qb-c}
 \end{gather}
with $c ={\rm  const}$.

\section[$q$-deformed Yang-Baxter algebra]{$\boldsymbol{q}$-deformed Yang--Baxter algebra}

The Yang--Baxter algebra underlying QIS and providing guarantee
for the integrability of a~discrete or lattice regularized quantum
system  was found to be given in the $q$-deformed  case by
a~general form  \cite{aa,aa1,aa2}
\begin{gather} [ S_q^ {+}, S_q^{-} ] =
 \big (\hat  M^+ \sin \big(2 \al S^3\big) +
\hat  M^-  \cos
\big( 2 \al S^3  \big) \big){1 \over \sin \al},\nonumber\\
[S^3,S_q^{\pm}] = \pm S_q^{\pm} , \qquad [ \hat M^\pm , \cdot]=0
.\label{ancAlg}
\end{gather}
Note that this is a novel type of deformed  algebra,
 which   apart from an
usual parameter $q$-de\-formation, also involves  additional
 operator-deformation by   central operators
 $\hat  M^\pm $.
Operator deformed quantum Yang--Baxter algebra (\ref{ancAlg}) is a
bialgebra and
 exhibits  noncommutative
Hopf algebra properties with the coproduct structures
\begin{gather}
 \De (S_q^+)= S_q^{+} \otimes
\hat c^-_{1}q^{-S^3 }  +
    \hat  c^-_{2} q^{S^3 }\otimes {S_q^+}, \qquad
   \De (S_q^-)= {S_q^-} \otimes
\hat c^+_{2}q^{-S^3 }   +
   \hat  c^+_{1}q^{S^3 }\otimes {S_q^-},
 \nonumber \\
\Delta( { S^3})= I \otimes { S^3}+S^3 \ot I, \qquad \Delta(\hat
c_a^{\pm })=\hat c_a^{\pm } \otimes \hat  c_a^{\pm }, \quad a=1,2,
\label{coprod}
\end{gather}
where  $\hat c^\pm_a$, $a=1,2 $ are related to central operators
as $\hat  M^\pm= \frac {1} {2 } (\hat c^+_1\hat c^-_2 \pm \hat
c^-_1\hat c^+_2 )$ and one  redef\/ines  $ q=e^{i \alpha}.$ Note
that  unlike  deforming
 parameter $q$,  deforming central
operators $\hat  c_a^{\pm } $ and hence $\hat M^\pm $
 have nontrivial coproduct.

The quantum Lax operator associated with this general
 YB algebra is
\begin{gather}
L_{\rm anc}(\xi) = \left( \begin{array}{cc}
  \xi{\hat c_1^+} e^{i \al S^3}+ \xi^{-1}{\hat c_1^-}  e^{-i \al S^3}&
2 \sin \al  S_q^-   \\
2 \sin \al  S_q^+    &  \xi{\hat c_2^+}e^{-i \al S^3}+
\xi^{-1}{\hat c_2^-}e^{i \al S^3}
          \end{array}   \right), \qquad
          \xi=e^{i \la}. \label{aL}
          \end{gather}

This ancestor model  \cite{aa,aa1,aa2} seems to be general enough
to generate all quantum integrable models with $(2 \times 2) $ Lax
operator
 associated
with the  trigonometric $R^{\rm trig}$ matrix
\begin{gather}
R^{11}_{11} = R^{22}_{22}= \sin (\la+\al), \qquad  R^{12}_{12} =
R^{21}_{21}= \sin  \la , \qquad R^{12}_{21} = R^{21}_{12}= \sin
\al.    \label{trm}
\end{gather}
Therefore  identifying the situations when the YB algebra
(\ref{ancAlg}) goes to dif\/ferent known algebras at dif\/ferent
values of the deforming central operators,
 we can construct from the corresponding reductions of the
Lax operator    (\ref{aL}) a series of quantum integrable models
solvable by the Bethe ansatz.

For example we may
 observe that in contrast to the well known
$q$-deformed algebras (\ref{qbos}),  (\ref{qs}),   algebra
(\ref{ancAlg})
 has
 two dif\/ferent terms in the RHS of its main
relation: the f\/irst one (sine-term)
  is similar to the quantum-spin algebra (\ref{qs}),
 while the second term  ( cosine-term)
 to the $q$-boson  (\ref{qbos}). Indeed
both these $q$-deformed
 algebras can be obtained  as   subalgebras of (\ref{ancAlg}).
Notice that $\hat M^+=1$, $\hat M^-=0  $ yields quantum algebra
$sl_q(2) $,
 while  $\hat M^+=-1$, $\hat M^-=0  $ gives   the noncompact algebra  $sl_q(1,1)$ as
def\/ined in (\ref{qs}). The  Lax operators of the  related
integrable systems can be obtained directly from the general form
(\ref{aL}), by using a compatible  choice $\hat c^+_1=\hat
c^+_2=1$, $\hat c^-_1=\hat c^-_2= \epsilon$, $\epsilon =\pm 1$ as
\begin{gather}
L_{qs}(\xi) = \left( \begin{array}{cc}
  \xi e^{i \al s^3}+ \epsilon \xi^{-1}  e^{-i \al s^3} &
2 \sin \al  s_q^-   \\
2 \sin \al  s_q^+    &  \xi e^{-i \al s^3}+ \epsilon \xi^{-1}e^{i
\al s^3}
          \end{array}   \right),  \label{Lqs}
\end{gather}
involving $q$-spin   belonging to $sl_q(2) $ and $sl_q(1,1) ,$ for
$\epsilon=\pm 1, $ respectively.

On the other hand for the particular values of the deforming
operators:
 $ \hat M^+= {\sin \al }$, $\hat M^-= i {\cos \al }$, one directly obtains the $q$-boson   (\ref{qbos})
as a subalgebra of (\ref{ancAlg}) with a  compatible  choice $\hat
c^+_1=\hat c^+_2=1$, $\hat c^-_1= -{iq }$, $\hat c^-_2=  {i \over
q}$. However note that such a choice of deforming operators, as
seen from (\ref{coprod}), seemingly does not respect the usual
coproduct structure and hence the corresponding algebra
representing $q$-boson exhibits  the known dif\/f\/iculty  in its
formulation of the Hopf algebra. We get the $q$-bosonic Lax
operator from the reduction of (\ref{aL}) as
\begin{gather}
L_{qb}(\xi) = \left( \begin{array}{cc}
  \xi e^{i \al N}-i \xi^{-1}  e^{-i \al (N-1)}&
\kappa b_q^\dag   \\
\kappa  b_q    &  \xi e^{-i \al N}+ i \xi^{-1}e^{i\al  (N-1) }
          \end{array}   \right),  \label{Lqb}
\end{gather}
with a direct mapping:
 \begin{gather}
  S_q^+=\frac {i\kappa} {\Omega} b_q, \qquad S_q^-
=\frac {i\kappa} {\Omega}b_q^\dagger,\qquad S^3=N, \qquad
\Omega=q-q^{-1}
 \label{sqbq}
 \end{gather}
and $\kappa ^2=\frac {q^2-q^{-2}} {4i} $. Interestingly if on the
other hand we choose
\begin{gather}
\hat c^+_1= \hat c^-_2=q^{\ha} \qquad  \mbox {with}  \qquad \hat
c^-_1= \hat c^+_2=0 \label{cqb}
\end{gather}
 or at least
 any one of $ \hat c^-_1$, $\hat c^+_2$
is $0$,  by representing
\begin{gather}
 S_q^+= \tilde \kappa b_q q^{\frac N 2},
 \qquad S_q^-=
\tilde \kappa q^{\frac N 2}b_q^\dagger,\qquad S^3=N, \qquad \tilde
\kappa ^2=\frac {i} { \Omega}, \label{sqbq1}
\end{gather}
 we   can get a direct realization to the $q$-bosonic
relations (\ref{qboson}) as a subalgebra of
 (\ref{ancAlg}).
 We
construct also the corresponding Lax operator from the related
reduction of (\ref{aL}). We see below that  such $q$-bosonic
models are intimately related to a series of interesting QIS
including
 the quantum lattice Liouville model.

\section[$q$-bosons in discrete QIS]{$\boldsymbol{q}$-bosons in discrete QIS}
We  identify and present here mostly  quantum integrable
$q$-bosonic models which can be  linked directly to Yang--Baxter
algebra and solved exactly through algebraic Bethe ansatz.

\subsection[$q$-Bose gas models]{$\boldsymbol{q}$-Bose gas models}

 i) A quantum integrable model involving $q$-bosons  constructed on a lattice
was proposed in \cite{bogolq}, which  can be  def\/ined simply
through  Lax operator (\ref{Lqs}) involving  $q$-spin operators
$(s^\pm_q, s^3) \in sl_q(2). $
 $q$-spin operators are in turn realized in single-mode  $q$-boson $b_q$, $b_q ^\dag $
using $q$-HP transformation (\ref{qHolst}). This lattice $q$-Bose
gas model associated with (\ref{Lqs}) and $ R^{\rm trig}$ exactly
satisf\/ies the QYBE by construction (since a particular case of
ancestor model (\ref{aL}))
 and therefore the related eigenvalue problem can
be solved   exactly through algebraic Bethe ansatz (ABA). In
analogy with the exact lattice model of nonlinear Schr\"odinger
(NLS) equation \cite{korlnls}
 the $q$-Bose gas model on the lattice can be
shown to have   localized Hamiltonian  \cite{bogolq}, as obtained
from the lattice NLS. However at the continuum limit the
$q$-bosonic mode turns into bosonic f\/ield and the model goes to
 the
same  NLS f\/ield model.

ii)  A $q$-harmonic oscillator model $ \ \ H_q=\frac {\hbar
\omega} {2}([N]_q + [N+1]_q) $ \ \
 was considered  in \cite{cheng02},
motivated by the noncommutative quantum f\/ield theory,
 in analogy with that of the harmonic
oscillator. Though this model does not belong to  quantum
integrable systems, it allows to calculate explicit result on
dif\/ferent thermodynamic properties like  free energy,
specif\/ic heat,
 entropy etc.

\subsection[Quantum lattice Liouville model as  $q$-bosonic model]{Quantum lattice Liouville model
as  $\boldsymbol{q}$-bosonic model}

Quantum integrable exact lattice version of the Liouville model
may be constructed as reduction of the above ancestor model, when
the deforming operators are chosen as (\ref{cqb}) as in  the case
of $q$-bosons and  generators of  algebra (\ref{ancAlg}) are
realized as
\begin{gather}
S^3=iu, \qquad  S_q^+=e^{-i\epsilon p}f(u), \qquad
S_q^-=f(u)e^{i\epsilon p},
 \label{liouville}
\end{gather}
where $f(u)^2=1-\epsilon ^2 q ^{-2iu+1} $. The Lax operator of
this lattice Liouville model can be obtained directly from
(\ref{aL}) and therefore its exact quantum integrability is
guaranteed.
 If one relates the parameter with
the lattice constant: $\epsilon=i\Delta$, then it is easy to show
that
 the model goes to the Liouville QFT model
at the continuum limit with the f\/ield Lax operator recovered
from the present discrete Liouville Lax operator.

On the other hand, it
 is interesting to observe that, the $q$-boson realization
(\ref{sqbq1}), using its mapping (\ref{qb-c}) through the
canonical variables turns unexpectedly to
 realization (\ref{liouville}) linked to  the lattice Liouville model.
Therefore we may conclude that
 the lattice
Liouville model is intimately related to and can be represented by
the $q$-bosonic model.

\subsection[Quantum derivative NLS through $q$-boson model]{Quantum derivative NLS through
 $\boldsymbol{q}$-boson model}

 Reduction of the ancestor model to  Lax operator
(\ref{Lqb}) expressed directly in $q$-bosons  gives another
quantum integrable deformed Bose gas model on the lattice.
$L$-operator (\ref{Lqb}) together with $ R^{\rm trig}$-matrix
(\ref{trm}) satisfy  the QYBE by construction and allow  exact
solution of this $q$-bo\-sonic
 model by algebraic AB (ABA).
 Using  mapping (\ref{qb-b}) the $q$-boson  def\/ined on a lattice
can be linked to a bosonic operator with commutation relation   $
[\psi_i , \psi^\dagger_j ] = {\hbar \delta_{ij} \over \Delta}$
 in the form
 \begin{gather} b_{qi} =
     \psi_i \left(\frac{[2N_i]_{q}}{2N_i \cos \al}\right)^{1/2}, \qquad N_i=
     \psi^\da_i\psi _i .
\label{bq-psi}
\end{gather}
It is interesting to note that
 at the continuum limit
$\Delta \to 0 $, when the lattice boson  $\psi _i $ produces
  the bosonic f\/ield $\psi _i \to \sqrt \Delta \psi (x) $,  the $q$-boson
through realization   (\ref{sqbq}) with the choice
\begin{gather} \hat c^+_1=\hat c^+_2=1, \qquad
\hat c^-_1= -\frac {\Delta} {4}{iq }  , \qquad \hat c^-_2= \frac
{\Delta} {4} {i \over  q}     \label{cdnls}
\end{gather}
 goes  to a quantum integrable  derivative NLS f\/ield model \cite{qdnls},
 given  by the equation
$  i \psi_t =  \psi_{xx}  -
  i  (  \psi^\dagger  \psi  )  \psi_x . $
The lattice Lax operator
 (\ref{Lqb}) with  mapping (\ref{bq-psi}), (\ref{cdnls})
 reduces at   $\Delta \to 0 $  to $ L=I+\De U^{\rm DNLS},$ where
\begin{gather*} U^{\rm DNLS}=
i \left( \begin{array}{cc}
  c \psi^\dagger \psi - {\lambda^2 \over 4 } &
 \sqrt c\, \lambda \psi^\dag   \\
\sqrt c\, \lambda \psi  & - c \psi^\dagger \psi + {\lambda^2 \over
4 }
          \end{array}   \right),
\end{gather*}
is the Lax operator of the DNLS f\/ield model \cite{qdnls}.
 Note that the quantum DNLS Hamiltonian
 in the $N$-particle sector   is  equivalent
 to the interacting Bose gas with {\it derivative} $\delta$-function
 potential~\cite{shirman}. We introduce in Section~7 a $q$-deformed anyon on the lattice and
subsequently construct an anyonic gas model interacting through
 $\delta'$-function, which is solvable by exact Bethe ansatz.

\subsection[Coupled  $q$-bosons and multi-component quantum integrable derivative NLS model]{Coupled
$\boldsymbol{q}$-bosons and multi-component quantum integrable\\
derivative NLS model}

Note that the Yang--Baxter  algebra (\ref{ancAlg}) is invariant
under the exchange of $\hat c^\pm_1 \leftrightarrow \hat c^\mp_2
$, since $\hat M^\pm $ do not change under such transformation.
However, this is not the case with the associated Lax operator
(\ref{aL}), which is transformed as $L(\xi) \to L^{-1}(\frac
{1}{\xi}) $.
 Therefore, using such a symmetry of  algebra (\ref{ancAlg})
 we can introduce another Lax
operator from  (\ref{Lqb}) with a similar but dif\/ferent
$q$-boson. Fusing these two Lax operators  we can construct a
novel   quantum integrable two-mode coupled $q$-bosonic lattice
model given by the Lax operator $L^{2qb}(\xi)=L^{qb}(\xi,
b_{q1})(L^{qb}(\frac {1}{\xi},
 b_{q2}))^{-1}$ with explicit expression  of its matrix elements as
\cite{hidden92}
\begin{gather}
 L^{2qb}_{11}=q^{-N_1+N_2}+i
\frac {\Delta} {4}\left(\frac {1} {\xi^2}q^{-(N_1+N_2+1)}+\xi^2
q^{N_1+N_2+1}\right) +\frac {\Delta ^2} {16}q^{N_1-N_2}+\kappa^2
b^\dagger _{q1}b_{q2}
,\nonumber\\
 L^{2qb}_{12}=\kappa \left(\frac {1} {\xi}\left(q^{-N_1}b^\dagger _{q2}
-i\frac {\Delta} {4}b^\dagger _{q1}q^{N_2+1} \right)+ {\xi}
(b^\dagger _{q1}q^{-N_2}) -i\frac {\Delta} {4}b^\dagger
_{q2}q^{N_1+1}  \right)
 ,\label{2qb}
\end{gather}
with $ L^{2qb}_{22}= (L^{2qb}_{11})^\dagger $ and $
L^{2qb}_{21}=( L^{2qb}_{12})^\dagger $. By combining $N^+ $ number
of $q$-boson of the f\/irst kind and  $N^- $ number from the
second kind  one can construct also a multi-mode generalization of
this
   $q$-bosonic model.
The mutually commuting conserved operators $C_k$, $k=0, \pm 1, \pm
2,\dots  $. in  the simplest case of two-mode  lattice $q$-bosonic
model may be generated by
\[ \tau (\xi)=tr \left(\prod_i^{N}L^{qb}(\xi, b_{q1i})
L^{qb}\left(\frac {1}{\xi}, b_{q2i}\right) \right) \] through
expansion  in spectral parameter: $\tau (\xi)=\sum\limits_kC_k
\xi^k. $ This
 long-range interacting lattice two-mode $q$-bosonic
model  is a new  quantum integrable model and
  should be solved  through algebraic
Bethe ansatz for  exact result.  The
 possibility of obtaining its local Hamiltonian
with  few nearest neighbor interactions by applying
 the method developed in \cite{korlnls}
and used for $q$-Bose gas in \cite{bogolq} is also an interesting
problem to explore.
 Considering
 $\Delta $ to be the lattice constant, we can go to
 the continuum limit  by taking $\De \to 0 $, where
 the two-mode  $q$-bosonic lattice
 Lax operator
(\ref{2qb})    turns into
 $L^{2qb}(\xi) \to I+i\Delta U^{\rm MTM}(\xi) $.
Interestingly,    Lax operator $U^{\rm MTM}(\xi) $ thus obtained
is the same  f\/ield Lax operator
  associated with
 the {\it bosonic}  massive Thirring model as well as
 with a two-component derivative
NLS model for  dif\/ferent choices for the  Hamiltonian. The
elements of $U^{\rm MTM}(\xi) $ are given by \cite{kulskly,qdnls}
\begin{gather*}
 U^{\rm MTM}_{11}= -U^{\rm MTM}_{22}=
\frac {1} {4}\left(\frac {1} {\xi^2}-\xi^2\right)+
\kappa_-\phi_1^\dagger \phi_1-\kappa_+\phi_2^\dagger \phi_2
, \nonumber\\
 U^{\rm MTM}_{12}=\big(  U^{\rm MTM}_{21}\big)^\dagger=
\xi \phi_1^\dagger +\frac {1} {\xi} \phi_2^\dagger.
\end{gather*}
We can evaluate the commuting conserved quantities of this
integrable f\/ield model  at the classical limit (putting
$\kappa_\pm  \to 1 $) using the Riccati equation:
\begin{gather*}
U_{12}\left({\nu \ov U_{12}}\right)_x +2U_{11} \nu +\nu
^2=U_{12}U_{21}.
\end{gather*}
We  solve this equation  by  expanding $\nu=\sum\limits_k C_{2k}
\xi^{2k} $ in spectral parameter at $\xi \to 0 $ and at $\xi \to
\infty $, obtaining the  conserved charges $C_{0}, C_2,C_4, \ldots
$ and  $C_{-0}, C_{-2},C_{-4}, \ldots $,
 respectively. In   explicit  form  they are given as
\begin{gather*}
C_0=|\phi_2|^2, \\ 
C_2=\phi_2^* \phi_1-4|\phi_2|^2|\phi_1|^2-2\phi_2^* \phi_{2x},\\
C_4=-C_{2x}-4\phi_2\phi_{1}^*|\phi_1|^2 +\frac {C_2}
{\phi_2^*}\left(\phi_{2x}^*-\frac 12 \phi^*_1\right) +(
|\phi_1|^2+ |\phi_2|^2 )(1-2C_2)+\phi_1 ^{*2}\frac {\phi_2}
{\phi_2^*}
 \end{gather*}
and
\begin{gather*}
C_{-0}=- |\phi_1|^2, \\ 
C_{-2}=\phi_1^* \phi_2-4|\phi_2|^2|\phi_1|^2+2\phi_1^* \phi_{1x},\\
C_{-4}=C_{-2x}+4\phi_1\phi_{2}^*|\phi_2|^2 -\frac {C_{-2}}
{\phi_1^*}\left(\phi_{1x}^*+\frac 12 \phi^*_2\right) -(
|\phi_1|^2+ |\phi_2|^2 )(1+2C_{-2} )-\phi_2 ^{*2}\frac {\phi_1}
{\phi_1^*}.
 \end{gather*}
It is interesting to observe  that  two dif\/ferent combinations
of the above  conserved operators of this integrable system can
generate two   important f\/ield models,
 which can  also be raised  to the quantum
level as quantum integrable systems solvable by the algebraic
Bethe ansatz.
 For example
the f\/ield
 Hamiltonian   given by
$H_{\rm MTM}=\int dx (C_2+C_{-2}) $ and the momentum
 $ \ P=\int dx (C_2-C_{-2}),  $  with canonical PB relations $
\{\phi_k(x), \phi_l^\dagger(y)\}_{\rm PB}=i \delta_{kl}\delta(x-y)
$
 would yield the relativistic
 massive Thirring model (MTM), rather its bosonic version.
On the other hand
 the higher Hamiltonian
$ H_{\rm 2DNLS}=\int dx (C_4-C_{-4}) $ would generate an
integrable 2-component DNLS  f\/ield model. At   quantum level it
would be an integrable QFT model,
 an exact  lattice regularized version of which
 we have constructed already
 through
two-mode $q$-bosonic  model.
  It is desirable
 to investigate this new  quantum  integrable model both at the discrete and
the continuum limit,  through exact Bethe ansatz formalism.

\subsection[$q$-boson and  Ablowitz-Ladik model]{$\boldsymbol{q}$-boson and  Ablowitz--Ladik model}

The celebrated  Ablowitz--Ladik (AL) model was
 discovered as a discretized version
of the NLS f\/ield model \cite{AL} much before the discovery of
$q$-boson.
 However   surprisingly one identif\/ies that the underlying YB
algebra of the quantum integrable AL model is actually given by
$q$-bosons.
 We observe f\/irst  that twisting
transformation
 $ R_{ij}^{kl} (\lambda ) \rightarrow e^{i\theta (j-k) }
 R_{ij}^{kl} (\lambda ),$
can give a new  $R$-matrix solution, where $\theta$ is some free
parameter. Using this twisted $R^{\rm trig}$-matrix one gets
a~$\theta $-deformation of the quantum algebra (\ref{ancAlg}), where
the  commutator  is deformed  as $pS_q^+S_q^--p^{-1}S_q^-S_q^+$,
$p=e^{i\theta} $, (similar to $p,q$-deformation)
\cite{aa,aa1,aa2}. As a result one gets a $\theta$-deformed
$q$-boson. For a special value $ \theta = - \alpha $ and with a
redef\/inition of $q$-boson as $ \tilde b_q \equiv b_q q^{-{N
\over 2}} \Omega^{1/2}
 $
and $\tilde b^{\dagger}_q \equiv q^{-{N \over 2}}  b_q^{\dagger
}\Omega^{1/2},
 $
one gets   another  form  of    $q$-boson
  with  algebra    $ q^2  \tilde b_q\tilde b^{\dagger}_q
 - \tilde b^{\dagger}_q \tilde b_q = q^2 - 1 $  as introduced by Macfarlane
\cite{qbose}. The Lax operator of the corresponding integrable
model as reduced from (\ref{aL}) is  given  simply~by
\begin{gather*} L_n =
 \left( \begin{array}{cc}
  { \xi^{-1} } & { \tilde b^{\dagger}_{qn} } \\ {\tilde b_{qn}}
 & \xi \end {array}
\right)
 ,
\qquad [ \tilde b_{qm} ,\tilde  b_{q n}^\dagger ] = \hbar ( 1-
\tilde b_{qn}^\dagger \tilde b_{qn} ) \delta_{m,n}
  \end{gather*}
with $\hbar = 1 - q^{-2}$, which  turns  out  to  be  the  lattice
model proposed by Ablowitz and  Ladik~\cite{AL} and its quantum
generalization discussed in~\cite{bogAL}.

We present another novel class of
 important  models involving $q$-bosons, e.g.\ $q$-deformed matter-radiation
models, in the next separate section.

 \section[Integrable  matter-radiation  models with $q$-bosons]{Integrable  matter-radiation
  models with $\boldsymbol{q}$-bosons}

Matter-radiation (MR) models represented by  atoms interacting
with  radiation are usually  described by    two-level spin
operators interacting with
 single mode boson. The well known and simplest
 models of this type are the Jaynes--Cummings (JC)  \cite{jc},
  and the Buck--Sukumar
(BS) model~\cite{bs}.
 The basic physics underlying a variety of  important
phenomena in interacting  MR systems, like those in quantum optics
induced by resonance interaction  between atom and a quantized
laser f\/ield,
 in cavity QED,
in trapped ion interacting with its center of mass motion
 irradiated by a laser beam
 \cite{trap}
 etc., seems to be nicely
 captured by such simple models.
Many theoretical predictions  based on these models, like
 vacuum Rabi splitting,
 Rabi oscillation  and its  quantum collapse and revival
etc.
 have been verif\/ied in maser and laser experiments.
 However, for describing physical situations more accurately
   generalizations of these basic models, like $q$-deformed BS
and JC  model \cite{qbs,qbs1,qjc,qjc1}, $q$-boson model
interacting with $q$-spins \cite{qbsBog} or its classical variant
\cite{qbsclBog}, trapped ion  (TI) with nonlinear coupling,
 multi-atom models \cite{vogel95,ntrap}
 etc.\ have been proposed.
However   most of the above generalizations, except a few
\cite{qbsBog},
  go beyond
  exact solutions and quantum integrability,
 especially for interacting multi-atom models.
Various $q$-bosonic models and quantum group symmetries were
discussed in~\cite{changq,changq1,changq2}.

We propose  new quantum integrable generalization of MR models
involving $q$-bosons and $q$-spins. A general approach for
constructing integrable MR model has been reported  in our earlier
work \cite{kunmr}. The physical signif\/icance of  $q$-deformed MR
models  in comparison with their undeformed counterparts is given
by their
  stronger
 nonlinear interactions between
 atomic excitations and the radiation mode,
as well as the possible presence of nonlocal and asymmetric
 interactions between atoms
due to  nontrivial coproduct      structure of $q$-spins. For
constructing these models in a unif\/ied way and for  ensuring
their quantum integrability we may  start from the integrable
 ancestor model (\ref {aL}) or more precisely by forming their Lax
operators  through  a~combination of (\ref  {Lqs})
   involving  $q$-spins and
(\ref{Lqb})  linked  to the $q$-boson. The general form of such
integrable $q$-deformed MR models may be given by the Lax operator
\begin{gather}
 L=L_{\rm anc}(\xi, S^\pm_q,S^3)
L_{qs}(\xi, s^\pm_q,s^3), \label{Lqmr}
\end{gather}
where $L_{\rm anc}(\xi, S^\pm_q,S^3) $  is given by (\ref {aL})
and
 $L_{qs}(\xi, s^\pm_q,s^3)
$ by (\ref {Lqs}) with $ \epsilon=+1$ . Therefore by construction
it would exactly satisfy the QYBE with
  ${\rm tr}\, (L)=\sum\limits_n C_n \xi^n $ def\/ining  the mutually commuting
 conserved operators  $C_n$, $n=0,1,2,\dots$  of this quantum integrable system.
 The Hamiltonian  of the  model given by $C_0 $
takes the form:
\begin{gather}
H_{q{\rm MR}}= H_d+  (s_q^+ S^-_q+ s^-_q S_q^+)\eta, \qquad
 H_d=c^+\cos \al (S^3-s^3) +ic^-\sin \al (S^3-s^3),
 \label{qimrh}
 \end{gather}
where $  \eta=4\sin^2 \al $, and  $c^\pm $ are related to the
values of the deforming operators appearing in the Lax operators.
This is a quantum integrable system with another commuting
conserved operator $C_2=S^3+s^3 $. Note that the quantum-spin
operator can be expressed through $N$-num\-ber of spin-$\ha $
operators using the coproduct structure def\/ined in the tensor
product of $N $
 vector spaces~as
\begin{gather}
s^\pm_q=\sum_j^{s} q^{-\sum\limits_{k<j}\sigma^3_k}
\sigma^\pm_jq^{\sum\limits_{l>j}\sigma^3_l}, \qquad s^3
=\sum_j^{s}\sigma^3_j.
 \label{sq-si}
 \end{gather}
This describes an asymmetric as well as nonlocal interaction
between the two-level atoms mediated by the $q$-bosonic radiation
mode.

Dif\/ferent reductions of this generalized $q$-deformed MR model
performed through possible  realizations of the generators
$S^\pm_q$, $S^3 $ of the ancestor algebra, as we have listed
above, can generate
 dif\/ferent physically relevant MR models in a unif\/ied way,  as we
describe below.

\subsection[Integrable  $q$-deformed Jaynes-Cummings model]{Integrable  $\boldsymbol{q}$-deformed Jaynes--Cummings model}
This model is constructed as a quantum integrable system of
interacting $q$-spins with $q$-boson representing atoms
interacting  nonlinearly
 with radiation as well
as among themselves. The  Lax operator of this integrable model
 can be constructed as a  reduction of
 (\ref{Lqmr}) by replacing the ancestor model by
(\ref{Lqb}) using the  $q$-bosonic realization (\ref{sqbq}).
 Consequently the Hamiltonian of the model can be
constructed from (\ref{qimrh}) as
\begin{gather}
H_{q{\rm JC}}= c \sin \al (N-s^3+\omega) + (b^\dagger_q s^+_q+
s^-_q b_q)\eta ,
 \label{qjch}\end{gather}
where $ c$, $\eta$, $\omega $ are  constant parameters (dependent
on $\alpha $),
  adjusted to simplify the expression. Notice that at  $q \to 1$ limit, when $
 s^\pm_q \to s^\pm, \  b_q \to b, $ the
$q$-deformed Jaynes--Cummings ($q$JC)  model~(\ref{qjch})
 goes to the integrable multi-atom Jaynes--Cummings    model (at $\alpha ^2 $
 order). In general this model
 describes  nonlinear and nonlocal interactions
 between atoms and the radiation mode,
since $s^\pm_q $ can be expressed through $N $-number of two-level
atoms in a nonlocal way  as~(\ref{sq-si})
 and the radiation represented by the
$q$-bosonic mode~$b_q$,~$b^\dagger_q $ is  expressed through
standard boson~$b$,~$b^\dagger $ in a nonlinear way through the
mapping
    (\ref{qb-b}).

Another similar multi-atom  $q$JC model with explicit interatomic
interactions may be given by def\/ining the Lax operator as
\begin{gather*}
L=L_{qb} \prod^N_{j} L^{(j)}_{xxz},
\end{gather*}
where $L^{(j)}_{xxz}$ is the Lax operator of anisotropic $xxz $
spin-$\ha $ chain, which represents here the $N$ interacting
two-level atoms. The corresponding Hamiltonian of this model would
include more complicated
  matter-radiation interactions as well as explicit
 atom-atom interactions.

Both the above  $q$JC  models are
 novel quantum integrable models solvable exactly by  algebraic Bethe
ansatz. Detailed analysis of these models with possible physical
importance  needs further  pursuing.

\subsection[Integrable  $q$-deformed Buck-Sukumar model]{Integrable  $\boldsymbol{q}$-deformed Buck--Sukumar model}

This quantum integrable model also describes matter-radiation
 interaction through
$q$-spins and a $q$-boson interacting in a stronger nonlinear way.
The idea of construction is  to start from the
  Lax operator  (\ref{Lqmr}) and take the realization of the
 ancestor model through generators of the noncompact quantum algebra
$su_q(1,1) $ as given by (\ref{Lqs}) for $\epsilon=-1 $.
Subsequently such generators are realized
 through $q$-boson via $q$-Holstein--Primakov transformation:
\[
s_q^+=\sqrt {[N]_q} b^\dag_q, \qquad s_q^-=b_q\sqrt{ [N]_q},
\qquad s_q^3=N+\ha  \] to represent the radiation mode.
 Consequently the Hamiltonian of this $q$-deformed Buck--Sukumar ($q$BS)
model which can be reduced from the general $q$MR model
(\ref{qimrh}) takes the form
\begin{gather*}
H_{q{\rm BS}}= c \sin \al (N-s^3+\omega)  + \eta \left
 (\sqrt {[N]_q}b^\dagger_q s^-_q+ s^+_q b_q \sqrt {[N]_q}\right) .
 \end{gather*}
The $q$-deformed  BS model gives an
 integrable
  version of an earlier  model \cite{qbs,qbs1}, when
$q$-spin operator is  replaced by $ {\vec \sigma} $-matrices by
taking spin-$ \ha$ representation.

 This novel
 quantum integrable $q$BS model can be exactly solved using Bethe ansatz,
which should produce physically interesting result generalizing
that of the  well known BS model, and therefore it deserves
detailed investigation. At $ q \to 0$, as    is clearly seen, the
$q$BS model
 goes to the integrable multi-atom BS
model.

Note that the quantum integrable model of  $q$-bosons  interacting
with $q$-spins proposed in \cite{qbsBog} is similar in spirit to
the present model, where however $q$-bosons were introduced not
directly but  as realization of $su_q(2) $ through $q$-HP
transformation.

\subsection[Integrable $q$-deformed trapped-ion model]{Integrable $\boldsymbol{q}$-deformed trapped-ion model}

Though this  model does not involve  $q$-bosons directly, we
present it here  since this novel quantum integrable MR model
belongs to the same trigonometric class  associated with the
quantum $R^{\rm trig} $-matrix and can be obtained  again  from
the $q$-deformed MR model (\ref{qimrh}) under suitable
realization.

Quantum Yang--Baxter algebra (\ref{ancAlg}) under  reduction
$\hat c^\pm_2=0,$ when both $\hat M^\pm=0,  $
 simp\-lif\/ies~to
\begin{gather*} [ S_q^ {+}, S_q^{-} ] =0
, \qquad  [S^3,S_q^{\pm}] = \pm S_q^{\pm} ,
\end{gather*}
allowing  realization  through canonical operators: $[x,p]=i , $
as
  \[ S_q^\pm=e^{\mp i x}, \qquad  S^3=p. \]
Note that the related reduction of (\ref{aL}) yields the Lax
operator of the quantum integrable relativistic Toda chain
\cite{rtoda,kuztoda,kuztoda1}. Interestingly the same realization
can reproduce from  the general MR Hamiltonian (\ref{qimrh})  a
 $q$-deformation of the    trapped ion ($q$TI) model
 given by
\begin{gather*}
H_{q{\rm TI}}= c_+ \cos \al (p-s^3+\omega) +c_- \sin \al
(p-s^3+\omega)  + \eta
 (e^{- i x} s^-_q+ e^{ i x} s^+_q )
 \end{gather*}
 with highly nonlinear coupling
between the atomic excitation and the vibration of the center of
mass motion of the trapped ion, described by displacement~$x $.
The system also has another  conserved operator $C_1=p+s^3 $.
 Usually for achieving exact
solution  such nonlinear oscillations are linearized through
several appro\-ximations like Dicke approximation, rotating wave
approximation etc.~\cite{trap}.
 The present model however is solvable in principle with full
exponential nonlinearity and without any approximation.

\section{Exact solution through algebraic Bethe ansatz}

 Almost all  $q$-deformed  models presented here
  similarly to their unif\/ied construction allow
  exact  ABA   solutions  also in a unif\/ied and
   vastly in a  model-independent way.

Note that from the local QYBE taking the tensor product
$T(\la)=\prod\limits_j L_j(\la) $ one can go to its
 global form:
\begin{gather}  R(\la-\mu)T(\la)\otimes T(\mu)=
(I \otimes T(\mu)) (T(\la)\otimes I)R(\la-\mu), \qquad j=1,2,
\dots ,N \label{gqybe}
\end{gather}
ref\/lecting the Hopf algebra structure of the underlying YB
algebra. Taking trace from both the sides of (\ref{gqybe})
 and def\/ining $ \tau(\la)={\rm tr }\, T(\la)$ we get $ [\tau(\la),\tau(\mu) ]=0$ and
expanding further in spectral parameter:
$\tau(\la)=\sum\limits_nC_n\la ^n $ derive f\/inally the quantum
integrability condition for the conserved operators: $ [C_n,C_m
]=0$.

ABA formalism  aims to solve exactly the eigenvalue problem for
all conserved operators simultaneously.  The diagonal entries
 $\tau (\la)=T_{11}(\la)
+T_{22}(\la)$
 produce the conserved operators, while the
 of\/f-diagonal elements $  T_{21}(\la)\equiv B(\la)$ and $
  T_{12}(\la) \equiv C(\la)$
   act like
creation and annihilation operators of
 pseudoparticles.  The $M$-particle  state is def\/ined as
 $|M\rangle_B=B(\la_1) \cdots B(\la_M)|0\rangle $ and the pseudovacuum
   $|0\rangle$  is def\/ined through
$ C(\la)|0\rangle=0$. The basic idea of algebraic BA \cite{aba} is
to f\/ind the eigenvalue solution: $\tau(\la)|M\rangle_B=\Lambda
(\la, \{ \la_a\})|M\rangle_B$, for which
 diagonal elements $ T_{ii}(\la)$, $i=1,2$
 are   pushed through the string of $B(\la_a) $'s toward  $|0\rangle$,
 using the commutation relations obtainable  from the
 QYBE (\ref{gqybe}).  Considering further the actions
$T_{11}(\la)|0\rangle=\alpha (\la)|0\rangle$,
$T_{22}(\la)|0\rangle=\beta (\la)|0\rangle$, one   arrives
f\/inally  at the eigenvalue expression
\begin{gather}
 \Lambda (\la,\{ \la_a\}) =
 \al(\la)\prod_{a=1}^M f(\la-\la_a)
+ \bet(\la)\prod_{a=1}^M f(\la_a-\la), \label{Lambda}
\end{gather}
where $f(\la)$ is def\/ined through the elements of the
 $R_{\rm trig}$-matrix as $
  {\sin (\la+\al) \ov \sin \la} .$ Expanding $ \Lambda (\la,\{ \la_a\}) $ in powers of $\la$ we  obtain
the eigenvalues for all conserved operators including the
Hamiltonian, where the rapidity  parameters $ \{ \la_a\} $
involved can be determined  from the Bethe equations
\begin{gather}
{ \al(\la_a) \ov \bet(\la_a)}=
 \prod_{b \not =a}{ f(\la_b-\la_a) \ov
 f(\la_a-\la_b)}
, \qquad a=1,2, \ldots , M, \label{be}
\end{gather}
which follow  in turn from the requirement of $|M\rangle_B$ to be
an eigenvector. Returning to our models we f\/ind that, the major
parts in  key algebraic BA relations (\ref{Lambda}) and
(\ref{be}),  described
  by   $R$-matrix elements $f(\la)$ is model-independent and hence
     same for all $q$-bosonic and
$q$-spin models  belonging to the  trigonometric class.

The only   model-dependent parts  in these equations, expressed
through $\al(\la)$ and $ \bet(\la) $    are
 determined  from the  Lax operator and therefore can be constructed in a
unif\/ied way starting from (\ref{aL}) as
\begin{gather}
 \al(\la)=c^+_1 \xi q^{m }+c^-_1\frac 1 \xi q^{-m },\qquad
  \bet(\la)=c^+_2 \xi q^{-m }+c^-_2\frac 1 \xi q^{m },
 \label{albet}\end{gather}
where $m=-\langle 0|S^3|0\rangle $ denotes the $q$-spin
projection.

 Since the $q$-bosonic or $q$-spin models
are obtained as various reductions of this ancestor model one can
obtain the corresponding exact result using the proper reduction
of~(\ref{albet}). Note however that for  some models like
relativistic Toda chain, $q$-deformed trapped ion model etc.\
since there is no easy pseudovacuum construction, the standard ABA
method is not applicable to them and more generalized functional
BA has to be implemented~\cite{rtoda}.

\section[$q$-deformed anyon and $\delta'$ anyon gas]{$\boldsymbol{q}$-deformed anyon and $\boldsymbol{\delta'}$ anyon gas}

We have found in Section~4.3, that a quantum integrable
$q$-bosonic model on a discrete lattice goes to an integrable
derivative NLS quantum f\/ield
 model involving
bosonic f\/ield opera\-tors~$\psi (x)$,~$\psi ^\dagger (x)  $. The
Hamiltonian of this integrable QFT model may be given by
  \begin{gather} H
=\int dx  \big( \psi^\dag_x \psi_x +i \kappa  \psi^\dag \psi
(\psi^\dag \psi_x- \psi^\dag_x \psi) \big), \label{hlle}
\end{gather}
which   at the $N$-particle sector can be shown to be equivalent
 to a derivative $\delta $-function Bose
gas model
  \begin{gather} {H}_N
=-\sum_k^N \partial^2_{x_k}+ i \kap \sum_{\langle k,l\rangle} \de
({x_k-x_l})\left(
 \partial_{x_k}+\partial_{x_l}\right)
\label{hdel1}
\end{gather}
which is
 exactly  solvable by the coordinate Bethe ansatz~\cite{shirman}.

Def\/ining a new notion  of $q$-anyon on the lattice we propose  a
similar
 model of 1d anyon gas interacting through  derivative $\delta $-function
and show that the model is also exactly solvable by coordinate BA.
Let us consider a lattice of $ N$-sites and let $q$-bosons:
 $b_{qj}$, $b_{qj}^\dagger$, $j=1,2, \ldots, N $ satisfying  commutation
relations (\ref{qbos}) for the same site: $[ b_{qj}, b_{qk}^ \dag
] = \delta_{jk}{ \cos \al (2N_j+1) \over \cos {\al } }$, etc.,
while   all operators commute at dif\/ferent sites with $j \neq k
$. We def\/ine  another set of nonlocal operators~as
\begin{gather}
 A_{qj}=e^{i \theta \sum\limits_{k=1}^JN_k} b_{qj}, \qquad A^\dagger _{qj}=
 b_{qj}^ \dag e^{-i \theta \sum\limits_{k=1}^JN_k}
\label{qa-qb}
\end{gather}
and easily check that at the same site these new operators behave
exactly like $q$-bosons with relations
\begin{gather}  [ A_{qj}, A_{qj}^ \dag ] = {
\cos \al (2N_j+1) \over \cos {\al } } ,\qquad [ A_{qj}, A_{qj}] =
0,\qquad \mbox{etc}.  \label{any0}
\end{gather}
However for dif\/ferent sites $j>l $, separated by any distance we
get an additional phase:
\begin{gather}
  A_{qj}A_{ql} =e^{i \theta} A_{ql}A_{qj}, \qquad   A^\dagger_{qj}
A^\dagger_{ql} =e^{i \theta} A^\dagger_{ql}A^\dagger_{qj} , \qquad
A_{qj} A^\dagger_{ql} =e^{-i \theta} A^\dagger_{ql}A_{qj},\qquad
\mbox{etc}. \label{any1}
\end{gather}
 $ \theta=0$ recovers obviously the $q$-bosonic case, while $\th=\pi $ gives
 anti-commutator at dif\/ferent sites similar to  fermions.
Coupling now  $q$-anyon to $q$-boson mapping (\ref{qa-qb}) with
that from $q$-boson to boson (\ref{qb-b}) and def\/ining the
boson- anyon transformation as
\begin{gather} A_{j}=e^{i \theta \sum\limits_1^JN_k} b_{j}, \qquad A^\dagger _{j}=
 b_{j}^ \dag e^{-i \theta \sum\limits_1^JN_k}, \qquad N_j= b_{j}^ \dagger b_{j}=
 A^\dagger _{j}A^\dagger _{j}
\label{a-b}
\end{gather}
  we can derive a similar
 mapping from
$q$-anyon to anyon as
\begin{gather}
  A_{qj}=A_jf(N_j), \qquad A^\dag_{qj}=f(N_j)A^\dag_j, \qquad  f(N)=
\left(\frac{[N]_{q}}{N }\right)^{1/2}, \qquad N=A^\dag A.
 \label{qa-a}
 \end{gather}
The commutation relations for the anyons $ A_j$, $A^\dag _j$ at
the same site are same as that of bo\-sons~(\ref{bos}), while at
dif\/ferent sites they
 are same as  (\ref{any1}).

 At the continuum limit $ \Delta \to 0$,  introducing a scaling through
$\sqrt \De $ the $q$-boson reduces to a~bosonic f\/ield $ \psi
(x)$:
\[b_{qi}  \to \sqrt \De \psi (x), \qquad N_j \to \Delta \psi^\dag (x) \psi (x). \]
 Consequently   the
$q$-anyon on the lattice is reduced to 1d anyon f\/ield $ \tilde
\psi (x)$:
\[ A_{qi} \to \sqrt \De \tilde  \psi (x), \qquad N_j \to \Delta \tilde
 \psi^\dag (x) \tilde  \psi (x), \]
where the anyonic f\/ield  satisf\/ies   the commutation
relations:
\begin{gather}
\tilde \psi^\dag (x_1) \tilde \psi^\dag (x_2)=e^{i \theta \ep
(x_1-x_2)}
\tilde \psi^\dag (x_2)\tilde \psi^\dag (x_1) , \nonumber \\
 \tilde \psi (x_1) \tilde \psi^\dag (x_2)=e^{-i \theta \ep (x_1-x_2)}
\tilde \psi^\dag (x_2)\tilde \psi (x_1)+ \de(x_1-x_2),\qquad
\mbox{etc.}, \label{cranyon}\end{gather}
 where
 \begin{gather}
  \ep (x-y)=\left\{ \begin{array}{rl} \pm 1 \quad & \mbox{for}\quad x >y, \quad x< y, \vspace{1mm}\\
0  \quad & \mbox{for} \quad x = y. \end{array}\right.\label{e}
 \end{gather}
From the introduction of the above discrete and 1d anyonic  and
$q$-anyonic operators it is clear that by replacing  $q$-bosons in
(\ref{Lqb}) by $q$-anyons  one can construct formally a
$q$-anyonic model. However unfortunately due to the nonlocal
commutation rules (\ref{any1}) the model turns into
a~nonultralocal model not satisfying the QYBE and hence becomes
nonintegrable. In a similar way if we consider the f\/ields in the
DNLS model  (\ref{hlle}) to be anyonic instead of bosonic, its
quantum integrability would be lost immediately.

However we f\/ind interestingly that, if instead of Bose gas we
consider (\ref{hdel1}) as the anyonic gas interacting through
 $\delta' $-function the model remains  exactly solvable
through coordinate Bethe ansatz,  though unlike
 the Bose gas  the anyonic  model is not a quantum integrable system with
mutually commuting higher conserved operators. Without giving the
details of this novel exactly solvable interacting anyon gas
model,
 which we reserve for a separate publication,
we mention only that the exact
 BA result of this model closely follow that of the
Bose gas  \cite{shirman} with a~redef\/inition of the coupling
constant $\kappa $ involving the anyon parameter $\theta $, as
happens
 also in the case of $\delta$-function anyon gas
\cite{delany,delany1}.

\section{Concluding remarks}
We have identif\/ied the appearance of $q$-bosons in quantum
integrable systems, exploring from the well known to  new models
in a unif\/ied way. Some  models, like Ablowitz--Ladik model,
  though well known, the underlying $q$-bosonic
connection of this model was not obvious and even unexpected. In
some
 cases involvement of
$q$-bosons  is more direct, like in the exact lattice version of
the quantum DNLS or the massive Thirring  model and in some
others, like in the $q$-Bose gas model of \cite{bogolq}, the
connection  to $q$-boson is only through  realization of $q$-spin
operators.
 Many  models   presented here, e.g.\ two-mode $q$-boson model and related
two-component DNLS quantum f\/ield model, $q$-deformed JC, BS and
TI models,
 as well as
 $\delta' $-function anyon gas,
   are new  quantum
integrable systems  with rich possibilities and deserve detailed
investigation. Since our objective here is to focus on various
integrable models and identify the role of $q$-bosons in them,
 we could not concentrate on any
individual  model  in detail, which we plan to do elsewhere.

\medskip

{\it Finally since this article is dedicated to the memory of
Vadim Kuznetsov, I would like to mention that   Vadim's interest
was closely linked to the present investigation, related to
quantum algebra and quantum integrable systems. Though
unfortunately I did not have any personal interaction with Vadim,
I remember having intense discussion with him through email
regarding our common interest  in formulating
 quantum relativistic
Toda chain model {\rm \cite{rtoda,kuztoda,kuztoda1}}, which is
intimately
  related  to the $q$-deformed
trapped ion model presented here.}

\pdfbookmark[1]{References}{ref}
\LastPageEnding
 \end{document}